\documentclass[review]{elsarticle}

\usepackage{lineno,hyperref}
\modulolinenumbers[5]

\journal{Materials Today Chemistry}

\begin{document}

\begin{frontmatter}

\title{Role of defects on carrier dynamics and transport mechanism in Bi$_2$Te$_3$ single crystals}

\author{Sonali Baral, Indu Rajput, Mukesh Kumar Dasoundhi, Devendra Kumar}
\author[mysecondaryaddress]{Archana Lakhani\corref{mycorrespondingauthor}}
\cortext[mycorrespondingauthor]{Corresponding author}
\ead{archnalakhani@gmail.com}
\address {UGC-DAE Consortium for Scientific Research, University Campus, Khandwa Road, Indore-452001}

\begin{abstract}
Defects play an important role in determining the type of carriers as well as on tunning the physical properties of layered materials. In this study, we have demonstrated that by varying the growth kinetics one can control the defects and can achieve electrons or holes dominated Bi$_2$Te$_3$ single crystals using modified Bridgman method. The corelation between structural defects and the type of dominant charge carriers in crystals are discussed using X-Ray diffraction and Hall resistivity. Electrons are found to be originating from Te vacancy type defects, while holes are manifested from predominant structural defects viz. Bi$_{Te}$ antisite defects or interstitial Te atoms. We observe that the alteration of charger carriers from electrons to holes have enhanced magnetoresistance (MR) from 103\% to 224\%. The enhancement in MR emerges from 2D multichannel quantum coherent conduction mechanism. 
\end{abstract}


\end{frontmatter}

\section{Introduction}
Topological insulator Bi$_2$Te$_3$ possess a single Dirac cone surface states perpendicular to any crystalline plane. Usually the surface states are observed through ARPES \cite{Chen2009} where the Dirac cone touches the top of the valence band. The surface carriers have been probed in Bi$_2$Te$_3$ through transport measurements via SdH oscillations \cite{Qu2010} \cite{barua2015} and weak antilocalization (WAL) \cite{shrestha2017}. But dominance of bulk carriers over surface carriers supresses the effect of surface states drastically, however the surface states can be observed via transport measurements in insulating Bi$_2$Te$_3$ samples whether dominant charge carriers are electrons (n) or holes (p). Bi$_2$Te$_3$ crystals can be formed with n or p-type dominant charge carriers. The suppression of surface carrier effects in transport results of metallic n-type Bi$_2$Te$_3$ have been reported by Qu et al. \cite{Qu2010} and a large Rashba splitting is also observed in metallic n-type Bi$_2$Te$_3$ \cite{holgado2020}. The n-type Bi$_2$Te$_3$ shows better thermoelectric performance compared to p-type Bi$_2$Te$_3$ \cite{writting2019}. Therefore, the prior information about the type of charge carriers would be advantageous in order to tune the desired properties of topological insulators.

Bi$_2$Te$_3$ crystalizes in rhombohedral structure of spacegroup \emph{R-3m} having five atomic layers (Te1-Bi-Te2-Bi-Te1) in a quintuple layer. These quintuple layers are bound with weak van der Waal forces and the atoms inside quintuple layer are tightly bound with each other through strong covalent bonds. Atomic imperfections such as vacancies or dislocations result in different types of native defects and consequently these native defects engender holes or electrons in the system. Te and Bi vacancies in Bi$_2$Te$_3$ provide two donors and three acceptors respectively, while Bi$_{Te}$ (substitution of Bi at Te atomic position), Te$_{Bi}$ (substitution of Te at Bi) antisite defects provide one acceptor and one donor respectively which eventually determine the type of charge carriers and carrier density. Thus, the p or n-type carriers in Bi$_2$Te$_3$ are controlled by the growth process. The phase diagram of Bi$_2$Te$_3$ clearly suggests that p-type carriers are dominant in Bi rich samples while n-type in Te rich samples \cite{writting2019} \cite{fleurial1988}. Earlier theoretical calculations have suggested the domination of n-type carriers in Te rich samples \cite{Miller1965} \cite{Pecheur1994}. However, there are contradicting experimental reports as well which have suggested that in Bi rich samples the most abundant defects are tellurium vacancies which induce n-type carriers in Bi$_2$Te$_3$ \cite{Netsou2020} \cite{Hashibon2011}. As an attempt to resolve this issue and understand the role of defects in determining the type of charge carriers, we have grown Bi$_2$Te$_3$ single crystals with two different protocols. The structural, morphological characterizations and transport measurements are performed on the same flakes of both the crystals. In this study, the role of growth conditions in determining the type of dominant charge carriers, whether p or n-type has been addressed with an inclusive comparision of literature and comprehensive magnetotranport measurements. We observe a large MR in p-type crystal compared to n-type crystal, which is due to swapping of classical to quantum transport behavior.

\begin{figure*}
	\centering
	\includegraphics[width=0.9\linewidth]{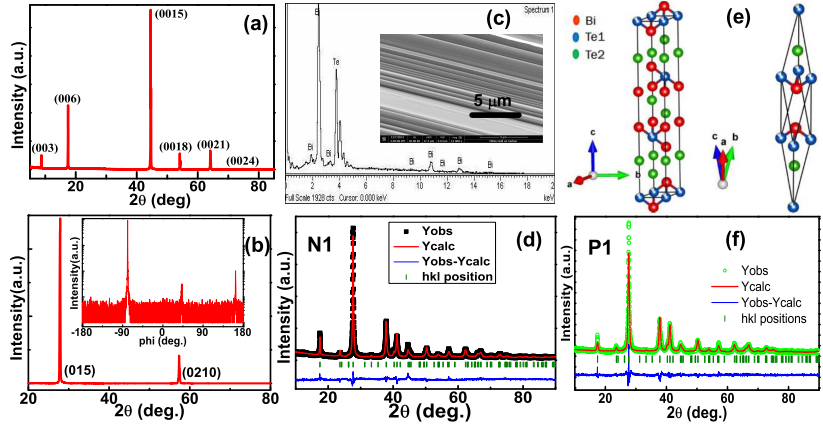}
	\caption{(a) Out-of-plane $\theta$-2$\theta$ XRD showing a highly oriented crystal grown along c-axis. (b) In-plane $\theta$-2$\theta$ XRD and phi scan (inset) of (015) plane. (c) EDX spectrum of same flake showing composition of Bi and Te is $\sim$2:3. Inset: SEM micrograph. (d) Rietveld refinement of powder XRD pattern for N1. (e) Hexagonal and rhombohedral lattice presentation using \emph{VESTA} software. (f) Rietveld refinement of XRD pattern of powdered sample of crystal P1.}
	\label{Fig.1}
\end{figure*}

\begin{table*}
	\centering
	\setlength{\tabcolsep}{2pt}
	\renewcommand{\arraystretch}{1}
	\begin {tabular} {|c|c|c|c|c|c|c|c|c|}
	\hline
	\textbf{Sample} &	\textbf{$\chi^2$} &	\textbf{a=b (\AA)} &	\textbf{c (\AA)} &	\textbf{V (\AA$^3$)} &	\textbf{Occ. Bi} &	\textbf{Occ. Te1} & 	\textbf{Occ. Te2} & \textbf{Strain (WH plot)} \\
	\hline
	N1 &	2.79 &	4.387(1) &	30.480(5) &	508.0(2) &	2 &	1.01(2) &	1.71(4) & 3.8(4)$\times$10$^{-4}$ \\
	\hline
	P1 &	2.13 &	4.388(1) &	30.488(6) &	508.1(3) &	2 &	0.89(2) &	2.45(7) & 3.9(1)$\times$10$^{-3}$ \\
	\hline
	\end {tabular}
	\caption{Structural parameters obtained from Rietveld refinement of both crystals N1 and P1 where a, b, c are lattice parameters; V volume; Occupancy (Occ.) of Bismuth (Bi), Tellurium of position 1 (Te1) and 2 (Te2). Strain calculated from WH plot of single crystal XRD.}
	\label{Table.1}
\end{table*}

\section{Experimental details}
We have grown Bi$_2$Te$_3$ crystals by modified Bridgman method. High purity Bi (99.9999\%) and Te (99.999\%) pieces were taken in the stochiometric ratio (2:3) and crushed into powder in an agate and mortar for several hours. The mixture was divided into two parts and was sealed in two separate evacuated quartz tubes in vacuum $\sim$10$^{-6}$ torr and melted for 24 hours at two different melting temperatures 950$^o$C (sample N1) and 900$^o$C (sample P1). The mixture was slowly cooled down to room temperature with a rate of 2$^o$C/hr. Since Bi$_2$Te$_3$ is a layered material having weak van der Waal forces between the quintuple layers, the crystals can be easily cleaved mechanically along ab-plane which is perpendicular to hexagonal c-axis. The out-of-plane and in-plane X-ray diffraction (XRD), phi-scan, $\omega$-scan XRD were performed using Bruker D8 advance and Bruker D8 discover X-ray diffractometer respectively with Cu-K$_\alpha$ source. Scanning electron microscopy (SEM) was carried out on FEI made Nova Nano field emission SEM (FESEM). Energy Dispersive X-ray (EDX) spectroscopy was done on same flake in JEOL JSM 5600 SEM instrument. Longitudinal and Hall resistivity measurements were performed on the same flakes using Physical property measurement system (PPMS) by varying the temperature from 300K to 6K and upto 9T magnetic field.

\section{Structural characterization and Transport results}

\subsection{Structural and morphological characterization}
Fig.1(a) displays the out-of-plane XRD of \{003\} planes signifying highly oriented nature of grown crystal N1 along c-axis. To confirm the single crystalline nature of crystal, in-plane XRD and $\phi$-scan of (015) plane were performed as shown in Fig.1(b) and in its inset respectively. The in-plane XRD shows the parallel \{015\} set of planes indicating the in-plane orientation of planes. The $\phi$-scan on (015) plane results in three peaks which are equally spaced with $\sim$120$^o$ demonstrating the three-fold symmetry of (015) plane. Both in-plane XRD and $\phi$-scan on (015) plane assert the single crystalline nature of the grown crystal. The SEM micrograph as shown in the inset of Fig.1(c) signifies the layered structure of the grown single crystal. Similar structural characterizations were carried out for crystal P1 which also confirms the single crystalline nature (not shown here). To confirm the phase purity, Rietveld refinement was carried out on powder XRD pattern using \emph{Fullprof Suite} software which verifies the single phase having spacegroup \emph{R-3m} (D$_{3d}$$^5$) and shows no impurity peaks. The refinement profiles are shown in Fig.1(d and f) for crystals N1 and P1 respectively. The schematic structures viz. hexagonal unit cell and rhombohedral primitive cells were constructed using \emph{VESTA} software and are presented in Fig.1(e). The refined parameters are tabulated in Table-1 for both the crystals. The lattice parameters of both crystals are same. The occupancies of crystal N1 obtained by Rietveld refinement Bi: Te1: Te2 = 2 : 1.01(2) : 1.71(4) are clearly indicating tellurium deficiency in Te2 position of crystal lattice. The occupancies obtained for crystal P1, Bi: Te1: Te2 = 2 : 0.89(2) : 2.45(7), indicate the formation of Te rich Bi$_2$Te$_3$ crystal. The DFT calculations suggest that Te1 vacancies have lowest formation energy than the Te2 vacancies and any other point defects that occurred in Bi$_2$Te$_3$ crystal \cite{Hashibon2011}. The deficiency of Te atoms in Te1 position is experimentally confirmed through STM for both stoichiometrically grown and Te rich Bi$_2$Te$_3$ crystals \cite{Netsou2020}. Usually antisite defects are formed in Te rich Bi$_2$Te$_3$ crystals. For further confirmation of antisite defects through type of charge carriers we have performed extensive Hall measurements.

\begin{figure}
	\centering
	\includegraphics[width=0.7\linewidth]{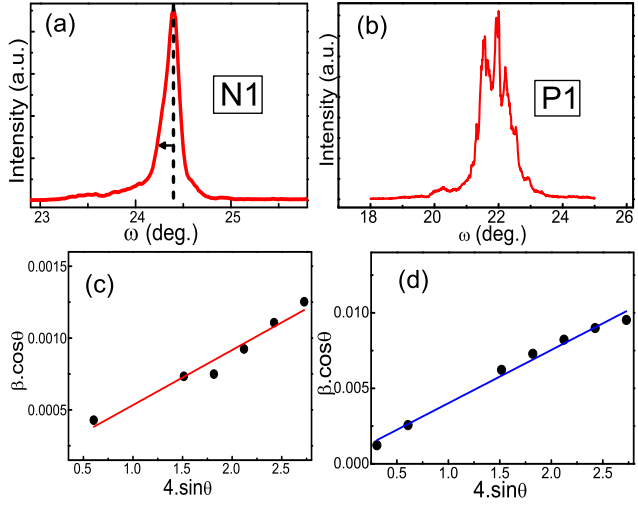}
	\caption{(a) $\omega$-scan of (0015) plane showing vacancy type defects encountered in crystal N1. (b) Rocking curve of (0015) plane showing low angle grain boundaries in crystal P1. (c and d) WH plot for N1 and P1 respectively.}
	\label{Fig.2}
\end{figure}

To confirm the crystalline quality $\omega$-scans are carried out on (0015) plane of both crystals as shown in Figs.2(a and b). Fig.2(a) shows a single peak indicating the high crystalline quality of the grown crystal. The left-hand side of rocking curve is broadened which signifies the presence of specific defects in the crystal. A theory of diffuse Huang scattering \cite{Huang1947} on point defects identifies this broadening as vacancy type of defects in a crystal. In vacancy type defects, lattice around the vacant atom gets expanded and hence the interplanar spacing, \emph{d} increases. As a result, the lower angle side of Bragg peak gets broadened while as in case of interstitial defects, the higher angle side gets broadened \cite{Baral2021}. The DFT calculations have suggested that V$_{Bi}$ has highest formation energy among all point defects in Bi$_2$Te$_3$ \cite{Hashibon2011}, thus the formation of Bi vacancy is least probable. Therefore, Tellurium vacancy (V$_{Te}$) is the only possibility, which is also indicated by Rietveld refinement of crystal N1. However, as shown in Fig.2(b), rocking curve of crystal P1 is deconvoluted into three peaks indicating the presence of low angle grain boundaries.

Williamson-Hall (WH) plot analysis was carried out for crystals N1 and P1 using equation: $\beta \cos\theta = k \frac{\lambda}{D} + 4 \epsilon\sin\theta$ where, $\beta$ is the broadening of diffraction line measured at half of its maximum intensity (FWHM), k is shape constant (0.9), $\epsilon$ is micro-strain, \emph{D} is crystallite size, $\theta$ is angle (in radian) and $\lambda$ is the wavelength (1.54 \AA). The WH plot is shown in Fig.2(c and d) for N1 and P1 respectively and the estimated strain values are tabulate in Table-1. The crystal P1 manifests higher strain compared to crystal N1.

\subsection{Longitudinal Resistance}

\begin{figure*}
	\centering
	\includegraphics[width=0.9\linewidth]{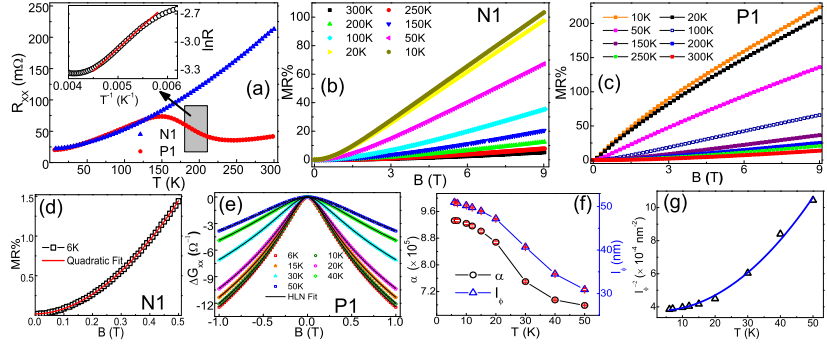}
	\caption{Temperature dependent resistance curves in the absence of magnetic field for Bi$_2$Te$_3$ crystals N1 (Blue triangle) and P1 (Red circle). Inset: Arrhenius plot for zero field resistance. Fit to the equation is shown with a red straight line for shaded regions. Isothermal magnetoresistance of crystals N1 (b) and P1 (c) respectively. (d) MR at 6K for crystal N1 (symbol) with quadratic fitting (red line). (e) MC at different temperatures for crystal P1 (symbols) with HLN fitting given by Eq.1 (black line). (f) HLN fitted parameters where $\alpha$= prefactor, $l_\phi$= coherence length. (g) $\frac{1}{l_\phi ^2}$ with temperature (symbols) and fitting to Eq.2 (blue line).}
	\label{Fig.3}
\end{figure*}

Temperature dependent resistance ($R_{xx}$) measurements are performed on both Bi$_2$Te$_3$ crystals N1 and P1 and shown in Fig.3(a) by blue triangle symbols and red circle symbols respectively. The $R_{xx}$(T) curve resembles a metallic behavior down to 6K for crystal N1. While $R_{xx}$(T) of crystal P1 first shows a metallic behavior down to 250K, an insulating behavior upto 150K and then metallic behavior down to 6K as shown in Fig.3(a). The insulating behavior is fitted from 212K to 183K with Arrhenius law expressed as: $R_{xx} = R_0 exp(-\Delta / k_B T)$, where $R_0$ is resistance at 0K, $\Delta$ is activation gap, k$_B$ is Boltzmann constant and T is temperature. The Arrhenius plot is shown in the inset of Fig.3(a) and $\Delta$ is found to be 41.6(2)meV which is comparable to Bi$_2$Te$_2$Se$_{1.05}$ ($\sim$45meV) \cite{RenBTS}, but larger compared to Bi$_2$Te$_3$ ($\sim$5meV) \cite{Lakhaniapl}, Bi$_2$Te$_2$Se ($\sim$23meV)\cite{RenBTS} and BiSbTe$_3$ ($\sim$4meV) \cite{indu2022} crystals and smaller compared to Sn doped Bi$_2$Te$_2$Se ($\sim$120 meV) \cite{RenBTS} and Bi$_2$Se$_2$Te ($\sim$100meV) \cite{BaoBST}. The residual resistivity ratio (RRR) of crystal N1 is found to be 9.6 which is comparable to other n-type Bi$_2$Te$_3$ crystals \cite{shrestha2017}.

Fig.3(b and c) display isothermal field dependent magnetoresistance (MR) for crystals N1 and P1 respectively. MR is found to be 103\% and 224\% at 10K and 9T for N1 and P1 respectively. The observed MR is large in crystal P1 compared to crystal N1. In crystal N1, throughout the measured temperature range, the MR shows a crossover from quadratic to linear dependence with magnetic field. The quadratic fitting at low magnetic field is shown in Fig.3(d) at 6K. This quadratic field dependence follows the semiclassical Boltzmann theory for metallic samples. However, crystal P1 shows quadratic to linear behavior of MR only at high temperatures, whereas below 50K cusp like behavior is found at low fields. The downward cusp like feature has been widely reported in topological insulators as weak antilocalization (WAL) effect which is well known as a quantum correction to magnetoconductance (MC). This quantum correction to MC has been explained by S. Hikami, A. I. Larkin, and Y. Nagaoka \cite{HLN1980} and the expression of simplified HLN equation is

\begin{equation}
	\label{eq3}
	\Delta G_{xx} (B) = - \frac{\alpha e^2}{\pi h} \bigg[\ln\bigg(\frac{B_\phi}{B}\bigg) - \Psi\bigg(\frac{1}{2} + \frac{B_\phi}{B}\bigg)\bigg],~ B_\phi = \frac{h}{8e\pi l_\phi^2}
\end{equation} 

where, $\Delta G_{xx}$(B) is change in conductance (G(B) - G(0)) as a function of magnetic field, $\alpha$ is a prefactor signifying number of 2D conduction channels, \emph{e} is electron charge, \emph{h} is Planck's constant, $\psi$ is Digamma function, B$_\phi$ is inelastic dephasing field and $l_\phi$ is coherence length. The above mentioned Eq.1 is a simpler form of Hikami-Larkin-Nagaoka (HLN) equation with a condition of $B, B_\phi \ll B_e, B_{so}$, where B$_{so}$ is spin-orbit and B$_e$ is elastic dephasing field. 

We have fitted MC curve with Eq.1 for temperatures from 6K to 50K as shown in Fig.3(e). The goodness of fit (Adj. R$^2$) is 0.99992 ($\sim$1) which signifies the presence of 2D WAL in the system. The obtained $\alpha$ and $l_\phi$ parameters are shown in Fig.3(f). Both $\alpha$ and $l_\phi$ increases with decreasing temperature. The reduction in $l_\phi$ indicates the weakening of WAL effect at higher temperatures. In TIs, the $\alpha$ value is 0.5 for single 2D conduction channel and 1 for two 2D parallel conduction channels. But, the prefactor $\alpha$ at 6K is $\sim$$9.3\times10^5$ which corresponds to $\sim$$1.86 \times 10^6$ number of transport channels, a value much larger than one or two 2D conduction channels. Thickness of crystal P1 signifies the existence of $\sim3.5\times10^5$ number of quintuple layers. In these layered materials, each quintuple layer behaves like a single 2D conduction channel \cite{Lakhaniapl} \cite{indu2022}. The comparable number of quintuple layers and number of 2D transport channels strongly suggests the presence of 2D multi-transport channels in the system. Consequently, the bulk of crystal P1 behaves like quantum coherent multi-transport channel system.

Further, $l_\phi$ is strongly temperature dependent. It is observed that the value of $l_\phi$ increases from 31nm to 51nm when temperature lowers from 50K to 6K.  Theoretically, $l_\phi \propto T^{-0.5}$ corresponds to Nyquist electron-electron (\emph{ee}) scattering decoherence mechanisms for 2D systems (TSS in TIs) and $l_\phi \propto T^{-0.75}$ corresponds to electron-phonon (\emph{ep}) interaction decoherence mechanism for 3D systems. But both these conditions are not satisfied in our case. Another temperature dependent parameter $l_\phi$ originates from combined effect of \emph{ee} and \emph{ep} interactions \cite{shrestha2017}\cite{Lakhaniapl} and is expressed as:
\begin{equation}
	\label{eq2}
	\frac{1}{l_\phi ^2 (T)} = \frac{1}{l_\phi ^2 (0)} + A_{ee} T^{p_1}  + A_{ep} T^{p_2}
\end{equation} 
where first term is zero temperature dephasing length which depends on sample geometry and defects, second and third terms represent the contributions from ee and ep interactions respectively which strongly depend on inelastic scattering time. For 2D case, $p_1$ = 1 and $p_2$ = 2. The $l_\phi$ is fitted using Eq.2 with $p_1$ = 1 and $p_2$ = 2, and fitting is shown in Fig.3(g) displaying a good match. The obtained parameters are $l_\phi$(0) = 51nm, $A_{ee}$ = -1.4$\times$10$^{-6}$nm$^{-1}$, $A_{ep}$ = 3$\times$10$^{-7}$nm$^{-2}$ which are comparable with other Bi$_2$Te$_3$ single crystals \cite{shrestha2017} \cite{Lakhaniapl}. The agreement of fitted curve with Eq.2 strongly suggests the involvement of both 2D ee and ep contributions in dephasing mechanisms. The good agreement of HLN equation in MC and 2D dephasing mechanisms from $l_\phi$ fitting strongly sugests that WAL is originated from 2D conduction channels.

\subsection{Hall resistivity}

\begin{figure}
	\centering
	\includegraphics[width=0.7\linewidth]{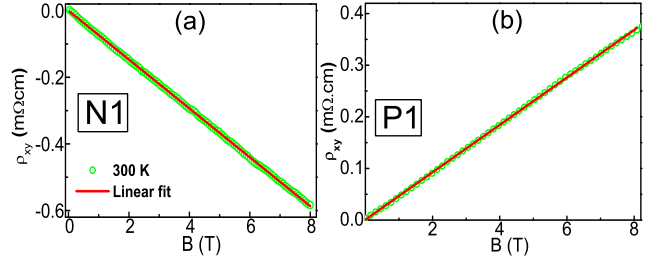}
	\caption{ Symmetrized Hall resistivity ($\rho$$_{xy}$) at 300K of crystals N1 (a) and P1 (b) respectively. Red solid line is the straight line fit to the data.}
	\label{Fig.4}
\end{figure}

In literature, theoretical reports have suggested both p-type \cite{Miller1965}\cite{Pecheur1994} and n-type charge carriers \cite{Netsou2020}\cite{Hashibon2011} for tellurium deficient Bi$_2$Te$_3$ samples. To confirm this in our case, Hall resistivity ($\rho$$_{xy}$) measurements were performed at 300K on both the crystals and graphs are shown in Fig.4(a and b). $\rho$$_{xy}$(B) demonstrates a negative slope indicating electrons as dominant charge carriers in crystal N1, while crystal P1 has holes as dominant carriers at room temperature. A straight line fit to $\rho$$_{xy}$ curve gives slope, R$_H$ and carrier density is calculated using a relation 1/eR$_H$ = $\sim$8.5$\times$10$^{18}$ cm$^{-3}$ and $\sim$1.67$\times$10$^{19}$ cm$^{-3}$ for N1 and P1 respectively.

\section{Discussions}
\subsection{Creation of n and p-type carriers:}
The Rietveld refinement results have shown the crystal N1 as Te deficient. In Bi$_2$Te$_3$, generally four types of defects occur viz. vacancies in Bi or Te site (V$_{Te}$ or V$_{Bi}$) and antisite defects (Bi$_{Te}$ or Te$_{Bi}$). In Bi rich or Te deficient Bi$_2$Te$_3$, most common defects encountered are V$_{Te}$ and Bi$_{Te}$. V$_{Te}$ acts as double donors and Bi$_{Te}$ acts as single acceptor. Thus, competition between these two defects will result in n or p-type carriers in crystal. In other words, even if both type of defects are present in the sample, the net carriers will be electrons only when the V$_{Te}$ defects dominate. From rocking curve and Rietveld refinement analysis, we have confirmed the presence of V$_{Te}$ defects in crystal N1. Thus, dominant defects V$_{Te}$ in Bi rich crystal N1 justifies the observation of n-type carriers from Hall measurement which agrees with earlier theoretical reports \cite{Hashibon2011} and STM results \cite{Netsou2020}.

The origin behind holes generation in Bi$_2$Te$_3$ is V$_{Bi}$ and Bi$_{Te}$ defects which provide three and one acceptors respectively. Theoretical calculations for Te rich Bi$_2$Te$_3$ have quantified that V$_{Bi}$ defects have highest formation energy and antisite defects Bi$_{Te}$ have lowest formation energy \cite{Hashibon2011}. Since Hall resistivity of crystal P1 confirms hole domination, there might be Bi$_{Te}$ antisite defects or interstitial Te atoms present in the sample. Rocking curve analysis yields low angle grain boundaries as shown in Fig.2(b) and strain is also found higher in crystal P1 than in crystal N1 as calculated from WH analysis (Fig.2(c and d) and Table-1). The presence of low angle grain boundaries and higher strain further supports the presence of Bi$_{Te}$ antisite defects or Te interstitials which induce holes in the crystal P1.

The formation of electrons in crystal N1 and holes in crystal P1 is associated with tellurium stoichiometry which can be further explained on the basis of crystal growth process. The latent heat of vaporization of Bi and Te is 50 Kcal/mole \cite{Fischer1966} and 27 Kcal/mole \cite{Brook1952} respectively. This suggests that the rate of evaporation of Te is more than Bi which can cause Te deficiency easily when heated near to its boiling point. During crystal growth process in case of crystal N1, stochiometric mixture of Bi and Te were melted at 950$^o$C while the boiling point of Te is 988$^o$C. In case of crystal P1, mixture was melted at 900$^o$C which is much below the boiling point of Te. Thus, different melting temperatures is the reason behind the formation of Te rich or deficient Bi$_2$Te$_3$ crystals which is further supported by literature. A report on Bi$_2$Te$_3$ single crystal has clearly shown the p-type charge carriers through Hall resistivity where Bi and Te were melted at 700-800$^o$C \cite{Huang2016} which is much smaller than boiling point of Te. Another report has shown n-type carriers where samples were melted above 1000$^o$C \cite{chen2013}. 

\begin{figure*}
	\centering
	\includegraphics[width=1\linewidth]{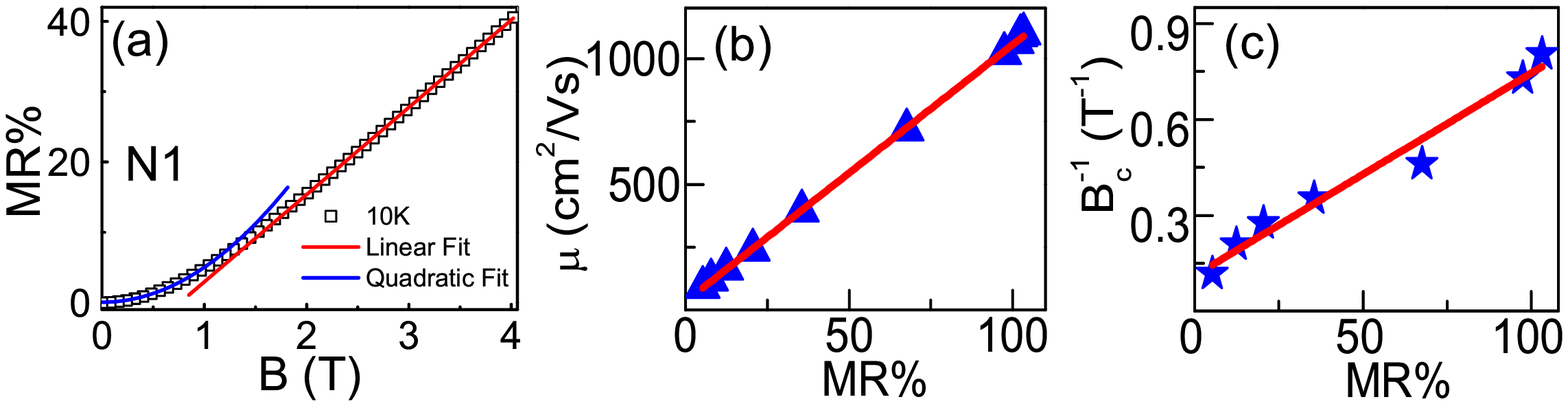}
	\caption{(a) MR upto 4T for crystal N1. Solid lines represents quadratic (blue) and linear (red) fit to the data. (b and c) MR\% vs. mobility ($\mu$) and inverse of crossover field (B$_c^{-1}$) graph for N1 respectively. Red solid line is straight line fit to the data.}
	\label{Fig.5}
\end{figure*}
 
\subsection{Comparision of magnetoresistance:}
In crystal N1, we observe quadratic MR at low fields which becomes linear at high fields as shown in Fig.5(a). The field at which MR behavior changes from quadratic to linear is known as crossover field (B$_c$). The linear MR (LMR) in N1 is directly proportional to mobility ($\mu$) and inversely proportional to B$_c$ as shown in Fig.5(b and c) respectively. This is well known Parrish and Littlewood (PL) criterion for the origin of LMR in chalcogenide systems \cite{Parish2003} \cite{Kumar2018}. Thus, the observed LMR in crystal N1, follows a classical behavior due to mobility fluctuations of charge carriers arising from defects present in N1. However, crystal P1 displays higher MR $\sim$224\% at 10K and 9T which does not tend to be linear and is not proportional to mobility as shown in Fig.6(a). Thus MR in crystal P1 does not follow the classical PL mechanism of mobility fluctuations instead we observe the quantum phenomena of WAL upto higher fields of 9T. The simplified HLN Eq.1 fits perfectly upto 4T as shown in the inset of Fig.6(b) with goodness of fit adj. R$^2$= 0.99994 at 10K, but shows a small deviation above 7.5T as shown by blue dashed line in Fig.6(b). In order to explore MR behavior at higher fields full HLN equation which includes spin-orbit coupling and elastic scattering terms must be considered and can be approximated as Eq.1 + $cB^2$ where \emph{c} is a constant \cite{Assaf2013}. We have fitted the MC in P1 upto 9T with Eq.1 + $cB^2$ (red solid line) which shows the best fit. The $\alpha$ value $\sim$1$\times$10$^{6}$ obtained from full HLN equation suggests the presence of $\sim$2$\times$10$^{6}$ 2D transport channels in the crystal P1 \cite{shrestha2017} \cite{Lakhaniapl} \cite{indu2022}. Thus the enhancement of MR in P1 is a consequence of quantum coherent phenomena where 2D multichannel conduction is taking place.

\begin{figure*}
	\centering
	\includegraphics[width=0.8\linewidth]{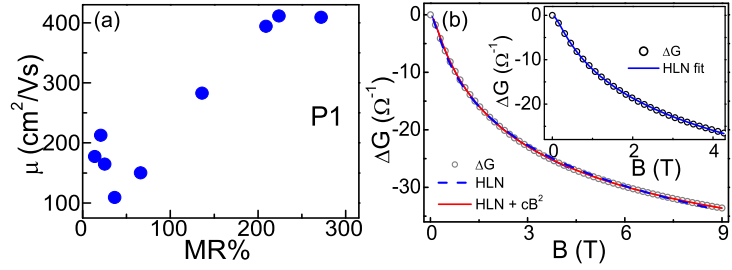}
	\caption{(a) MR vs. $\mu$ of crystal P1. (b) MC curve of P1 at 10K data upto 9T. Blue dashed line is Eq.1 fit and red solid line is Eq.1+$cB^2$ fit to the data. Inset: MC curve with HLN Eq.1 fit to the data (blue line).}
	\label{Fig.6}
\end{figure*}

\section{Conclusion}
In conclusion, two stochiometrically grown Bi$_2$Te$_3$ single crystals melted at two different temperatures have revealed two different type of dominant charge carriers in Hall resistivity. Bi rich Bi$_2$Te$_3$ crystal (N1) has shown electrons, while Te rich crystal (P1) has shown holes. The n-type sample is formed due to tellurium vacancies which occures because of high evaporation tendency of Te during crystal growth and the p-type crystal is formed due to Bi$_{Te}$ antisite defects or interstitial Te atoms. Low field MR unveils a classical behavior in metallic crystal N1, while semiconducting crystal P1 reveals WAL effect, a quantum correction to MC which is well explained by HLN theory suggesting the presence of 2D coherent multichannel conduction mechanism in this system. The investigation reported here, has revealed an effective way to tune the charge carriers from electrons to holes and thus, the transport behavior inherently changes from classical to quantum mechanism which further enhances the MR remarkably. Subsequently tuning the desired transport behavior paves a constructive way for employing these topological materials for applications in quantum information technology.

\section{Acknowledgments}
We thank V. R. Reddy and A. Gome for HRXRD measurements; M. Gupta and L. Behera for XRD measurements; R. Venkatesh and S. Potdar for SEM measurements; V. K. Ahire for EDX measurements.


\begin{thebibliography}{}
	\bibitem [1] {Chen2009} {Y. L. Chen, J. G. Analytis, J.-H. Chu, Z. K. Liu, S.-K. Mo, X. L. Qi, H. J. Zhang,	D. H. Lu, X. Dai, Z. Fang, S. C. Zhang, I. R. Fisher, Z. Hussain, Z.-X. Shen, Experimental realization of a three-dimensional topological insulator, Bi$_2$Te$_3$, Science \textbf{325}, 178 (2009). DOI: 10.1126/science.1173034}
	
	\bibitem [2] {Qu2010} { D. X. Qu, Y. S. Hor, J. Xiong, R. J. Cava, and N. P. Ong, Quantum oscillations and Hall anomaly of surface states in the topological insulator Bi$_2$Te$_3$, Science (80). \textbf{329}, 821 (2010). DOI: 10.1126/science.1189792}
	
	\bibitem [3] {barua2015} {S. Barua, K. P. Rajeev, and A. K. Gupta, Evidence for topological surface states in metallic single crystals of Bi$_2$Te$_3$, J. Phys. Condens. Matter \textbf{27}, 015601 (2015). Doi:10.1088/0953-8984/27/1/015601}
	
	\bibitem [4] {shrestha2017} {K. Shrestha, M. Chou, D. Graf, H. D. Yang, B. Lorenz, and C. W. Chu, Extremely large nonsaturating magnetoresistance and ultrahigh mobility due to topological surface states in the metallic Bi$_2$Te$_3$ topological insulator, Phys. Rev. B \textbf{95}, 195113 (2017). DOI: 10.1103/PhysRevB.95.195113}
	
	\bibitem [5] {holgado2020} { D. P. A. Holgado, K. Bolaños, S. De Castro, H. S. A. Monteiro, F. S. Pena, A. K. Okazaki, C. I. Fornari, P. H. O. Rappl, E. Abramof, D. A. W. Soares, and M. L. Peres, Shubnikov-de Haas oscillations and Rashba splitting in Bi$_2$Te$_3$ epitaxial Film, Appl. Phys. Lett. \textbf{117}, 102108 (2020). Doi: 10.1063/5.0019081}
	
	\bibitem [6] {writting2019} {I. T. Witting, T. C. Chasapis, F. Ricci, M. Peters, A. N. Heinz, G. Hautier, and G. J. Snyder, The thermoelectric properties of Bismuth Telluride, Adv. Electron. Mater. \textbf{5}, 1800904 (2019). https://doi.org/10.1002/aelm.201800904}
	
	\bibitem[7]{fleurial1988} {J. P. Fleurial, L. Gailliard, R. Triboulet, H. Scherrer, and S. Scherrer, Thermal properties of high quality single crystals of Bishmuth Telluride- Part I: Experimental characterization, J. Phys. Chem. Solids \textbf{49}, 1237 (1988). https://doi.org/10.1016/0022-3697(88)90182-5}
	
	\bibitem [8] {Miller1965} {G. R. Miller and C. Y. Li, Evidence for the existence of antistructure defects in Bismuth Telluride by density measurements, J. Phys. Chem. Solids \textbf{26}, 173 (1965). https://doi.org/10.1016/0022-3697(65)90084-3}
	
	\bibitem[9]{Pecheur1994} {P. Pecheur and G. Toussaint, Tight-binding studies of crystal stability and defects in Bi$_2$Te$_3$, J. Phys. Chem. Solids \textbf{55}, 327 (1994). https://doi.org/10.1016/0022-3697(94)90229-1}
	
	\bibitem[10]{Netsou2020} {A. M. Netsou, D. A. Muzychenko, H. Dausy, T. Chen, F. Song, K. Schouteden, M. J. Van Bael, and C. Van Haesendonck, Identifying native point defects in the topological insulator Bi$_2$Te$_3$, ACS Nano \textbf{14}, 13172 (2020). https://dx.doi.org/10.1021/acsnano.0c04861}
	
	\bibitem[11]{Hashibon2011} {A. Hashibon and C. Elsässer, First-principles Ddensity functional theory study of native point defects in Bi$_2$Te$_3$, Phys. Rev. B - Condens. Matter Mater. Phys. \textbf{84}, 144117 (2011). DOI: 10.1103/PhysRevB.84.144117}
	
	\bibitem[12]{Huang1947} {K. Huang, X-Ray Reflexions from dilute Solid Solutions., Proc. R. Soc. Lond. A. Math. Phys. Sci. \textbf{190}, 102 (1947). https://doi.org/10.1098/rspa.1947.0064}
	
	\bibitem[13]{Baral2021} {S. Baral, M. K. Dasoundhi, I. Rajput, D. Kumar, and A. Lakhani, Large linear magnetoresistance and evidence of degeneracy lifting of valence Bands in rhombohedral phase of topological crystalline insulator SnTe, Phys. Status Solidi – Rapid Res. Lett. \textbf{16}, 2100542 (2022). DOI: 10.1002/pssr.202100542}
	
	\bibitem[14]{RenBTS} {Zhi Ren, A. A. Taskin, Satoshi Sasaki, Kouji Segawa, and Yoichi Ando, Fermi level tuning and a large activation gap achieved in the topological insulator Bi$_2$Te$_2$Se by Sn doping, Phys. Rev. B \textbf{85}, 155301 (2012). https://doi.org/10.1103/PhysRevB.85.155301}
	
	\bibitem[15]{Lakhaniapl} {A. Lakhani, D. Kumar, Observation of multichannel quantum coherent transport and electron-electron interaction in Bi$_2$Te$_3$ single crystal, Appl. Phys. Lett. \textbf{114}, 182101 (2019).https://doi.org/10.1063/1.5089536}
	
	\bibitem[16]{indu2022} {I. Rajput, S. Baral, M. K. Dasoundhi,  D. Kumar, and A. Lakhani, Quantum coherent transport and electron–electron interaction in BiSbTe$_3$ single crystals, Materials Today Communications \textbf{33}, 104537 (2022). https://doi.org/10.1016/j.mtcomm.2022.104537}
	
	\bibitem[17]{BaoBST} {Lihong Bao et al., Weak anti-localization and quantum oscillations of surface states in topological insulator Bi$_2$Se$_2$Te, Sci. Rep. \textbf{2}, 726 (2012). https://www.nature.com/articles/srep00726}
	
	\bibitem[18]{HLN1980} {S. Hikami, A. I. Larkin, and Y. Nagaoka, Spin-Orbit interaction and magnetoresistance in the two dimensional random system, Prog. Theor. Phys. \textbf{63}, 707 (1980). https://doi.org/10.1143/PTP.63.707}
	
	\bibitem[19]{Fischer1966} {A. K. Fischer, Vapor pressure of Bismuth, J. Chem. Phys. \textbf{45}, 375 (1966). http://dx.doi.org/10.1063/1.1727337}
	
	\bibitem[20]{Brook1952} {L. S. Brooks, The vapor pressures of Tellurium and Selenium, J. Am. Chem. Soc. \textbf{74}, 227 (1952). https://doi.org/10.1021/ja01121a059}
	
	\bibitem[21]{Huang2016} {S. M. Huang, S. Y. Lin, J. F. Chen, C. K. Lee, S. H. Yu, M. M. C. Chou, C. M. Cheng, and H. D. Yang, Shubnikov-de Haas oscillation of Bi$_2$Te$_3$ topological insulators with cm-Scale uniformity, J. Phys. D. Appl. Phys. \textbf{49}, 255303 (2016). Doi: 10.1088/0022-3727/49/25/255303}
	
	\bibitem[22]{chen2013} {T. Chen, J. Han, Z. Li, F. Song, B. Zhao, X. Wang, B. Wang, J. Wan, M. Han, R. Zhang, and G. Wang, Shubnikov de Haas quantum oscillation of the surface states in the metallic Bismuth Telluride sheets, Eur. Phys. J. D \textbf{67}, 75 (2013).}
	
	\bibitem[23]{Parish2003} {M. M. Parish and P. B. Littlewood, Non-saturating magnetoresistance in heavily disordered semiconductors, Nature \textbf{426}, 162 (2003). https://www.nature.com/articles/nature02073}
	
	\bibitem[24]{Kumar2018} {D. Kumar and A. Lakhani, Large linear magnetoresistance from neutral defects in Bi$_2$Se$_3$ single crystal, Phys. Status Solidi RRL,\textbf{12}, 1800088 (2018). DOI: 10.1002/pssr.201800088}
	
	\bibitem[25]{Assaf2013} {B. A. Assaf, T. Cardinal, P. Wei, F. Katmis, J. S. Moodera, and D. Heiman, Linear magnetoresistance in topological insulator thin films: Quantum phase coherence effects at high temperatures, Appl. Phys. Lett. \textbf{102}, 012102 (2013). https://doi.org/10.1063/1.4773207}
\end{thebibliography}
\end{document}